\documentclass[reprint,superscriptaddress,amsmath,amssymb,pra,noeprint]{revtex4-2}

\usepackage{graphicx}
\usepackage{dcolumn}
\usepackage{bm}
\usepackage{amsmath,amsthm,amssymb,physics,xcolor,enumerate,tabularx,makecell,mathrsfs,bbm,stackrel,mathtools,multirow,bm}

\usepackage[normalem]{ulem}

\usepackage[breaklinks=true,colorlinks=true,linkcolor=teal,urlcolor=teal,citecolor=teal]{hyperref}

\newcommand{\id}{\mathbbm{1}}

\begin{document}

\title{Fisher information susceptibility for multiparameter quantum estimation}

\author{Francesco Albarelli}
\affiliation{Scuola Normale Superiore, I-56126 Pisa, Italy}

\author{Ilaria Gianani}
\affiliation{%
Dipartimento di Scienze, Universit\`a degli Studi Roma Tre, Via della Vasca Navale 84, 00146 Rome, Italy
}%

\author{Marco G. Genoni}
\affiliation{Department of Physics ``A. Pontremoli", Universit\`a degli Studi di Milano, I-20133 Milano, Italy}%

\author{Marco Barbieri }
\affiliation{%
Dipartimento di Scienze, Universit\`a degli Studi Roma Tre, Via della Vasca Navale 84, 00146 Rome, Italy
}%
\affiliation{%
Istituto Nazionale di Ottica - CNR, Largo E. Fermi 6, 50126 Florence, Italy
}%

\date{\today} 

\begin{abstract}
Noise affects the performance of quantum technologies, hence the importance of elaborating operative figures of merit that can capture its impact in exact terms.
In quantum metrology, the introduction of the Fisher information measurement noise susceptibility now allows to quantify the robustness of measurement for single-parameter estimation.
Here we extend this notion to the multiparameter quantum estimation scenario.
We provide its mathematical definition in the form of a semidefinite program.
Although a closed formula could not be found, we further derive an upper and a lower bound to the susceptibility.
We then apply these techniques to two paradigmatic examples of multiparameter estimation: the joint estimation of phase and phase-diffusion and the estimation of the different parameters describing the incoherent mixture of optical point sources.
Our figure of merit provides clear indications on conditions allowing or hampering robustness of multiparameter measurements.
\end{abstract}

\maketitle

\section{Introduction}

The characterization of any physical system relies on the precise estimation of the physical parameters entering its description. 
This task has fundamental importance from both a purely theoretical and a technological perspective. 
It can be thought of as a multiparameter estimation problem to be addressed with the tools of quantum metrology, which aims to derive the ultimate bounds on the precision of parameter estimation in the quantum realm and to identify possible enhancements by exploiting quantum resources, such as entanglement and squeezing~\cite{Giovannetti1330,Giovannetti2006,Giovannetti2011, PhysRevD.23.1693, Paris2009, Demkowicz-Dobrzanski2015a,Degen2016, Pirandola2018}.
While most of the paradigmatic results have been obtained in the context of single-parameter estimation, such as phase and frequency, much attention has also been devoted to the case of multiparameter quantum metrology~\cite{MultiPerspective,Szczykulska2016,Liu_2020,DemkowiczDobrzanski2020}. 
The field encompasses problems such as the simultaneous estimation of multiple phases \cite{Genoni2013b, Bradshaw2017, Humphreys2013, Gagatsos2016a, Knott2016, Pezze2017}, of unitary and noise parameters~\cite{knysh2013estimation,Vidrighin14,Altorio2015,Szczykulska_2017,Roccia_2018} and of the parameters describing illumination from incoherent optical point sources~\cite{TsangPRX2016,Chrostowski2017,Rehacek2017,Rehacek18PRA,Napoli2019,Fiderer2021,Ansari2021}.
The practical relevance of these problems justifies the introduction of the new challenges linked to measurement incompatibility when searching for the optimal conditions.
In fact, the optimal measurements to estimate distinct parameters individually may not commute.
This captures a characteristic aspect of quantum mechanics under a different light~\cite{Ragy2016}, and provides means to introduce a notion of ``quantumness'' for quantum statistical models~\cite{Carollo2019,Razavian2020,Candeloro2021}.
In addition, while collective measurements over multiple identical copies bring no advantage for single-parameter estimation, this strategy is instead crucial to achieving the ultimate precision bounds in the multiple-parameter case~\cite{MultiPerspective,Roccia_2018,Belliardo2021}.

Undesired interaction with the environment and, more in general, the noise affecting a quantum system is as detrimental for quantum metrology, as for all quantum technologies.
The relative simplicity of the single-parameter scenario has allowed the discovery of the ultimate bounds on the achievable precision in the presence of uncorrelated noise, and of the corresponding optimal protocols~\cite{Fujiwara2008,EscherNatPhys,Demkowicz-Dobrzanski2012,Chaves2013,Brask2015,Smirne2015a,Albarelli2018Quantum,Zhou2018,Zhou2019e,Rossi2020,Zhou2020,Kurdzialek2023a,Kurdzialek2024}.
Progress in the multiparameter scenario has been hindered by the strictly quantum problem of measurement incompatibility, as well as by multivariate nature of the underlying statistical problem.
Despite these challenges, generalization of single-parameter bounds have appeared~\cite{Katariya2020b,Albarelli2022,Hayashi2024}, as well as optimal strategies when the noise is correctable~\cite{goreckiOptimalProbesErrorcorrection2020}.
The theory of asymptotic multiparameter quantum estimation, however, is far from being complete.

In all these approaches, the effect of noise is associated either to an imperfect state preparation or to a noisy evolution realising the parameter encoding.
Only recently, the impact of noise at the level of the measurement stage has been addressed in different contexts~\cite{LenNatComm2022,Menos,zhou2023optimal}.
In particular, in Ref.~\cite{Menos}, the authors have investigated the effect of an imperfect realization of a given quantum measurement on the estimation precision; in order to properly quantify this impact, a new figure of merit, dubbed Fisher Information Measurement Noise Susceptibility (FI MeNoS), was introduced.
In this article we extend their treatment to the multiparameter scenario, by introducing the Fisher Information Matrix-Suited Measurement Noise Susceptibility (FI MaS MeNoS).
Besides giving the corresponding mathematical definition and deriving general formulas that will allow its evaluation, we will focus on two paradigmatic examples: the joint estimation of phase and phase-diffusion with a qubit system, and the estimation of the separation of two close incoherent sources along with other instrumental parameters.
Although we did not succeed at obtaining closed expressions, we show that the problem naturally leads to a numerical-friendly formulation as a semidefinite program.

The article is structured as follows: in Sec.~\ref{s:singleMenos} we review the single-parameter FI MeNos introduced in Ref.~\cite{Menos}, while in Sec.~\ref{s:multiMenos} we define the multi-parameter FI MaS MeNos, along with the derivation of upper and lower bounds.
In Sec.~\ref{s:examples} we apply our definition to the two examples mentioned above: i) the estimation of phase and phase-diffusion encoded in single-qubit systems, evaluating FI MaS MeNoS for separable and collective measurement strategies; ii) the estimation of the separation between two sources using the optimal measurement that, in ideal conditions, overcomes Rayleigh's curse.
This turns in to a multiparameter problem when we realise that we must also determine the centroid of the emission as well as the ratio of the two intensities~\cite{Hradil2017,Ansari2021}.

\section{Review of the FI MeNoS for the single-parameter case}\label{s:singleMenos}

We start by revising the definition and the most important properties of the FI MeNoS introduced in Ref.~\cite{Menos} by Kurdziałek and Demkowicz-Dobrzanski, recalling the basic ingredients of single-parameter quantum estimation theory. 

We consider a quantum statistical model, i.e. a family of quantum states $\rho_{\theta}$ labelled by a continuous parameter $\theta$ that one wants to estimate.
If one performs a measurement, mathematically described by a positive operator-valued measure (POVM) $M$ with elements $\{M_\alpha\}_{\alpha=1,\dots,O}$, the quantum statistical model is reduced to the classical statistical model $p(\alpha | \theta)$ that corresponds to the conditional probability of obtaining the measurement outcome $\alpha=1,\dots, O$ given the value of the parameter $\theta$, i.e. $p(\alpha|\theta) = \text{Tr}[\rho_{\theta} M_\alpha]$, according to the Born rule.
The variance of any unbiased estimator of the parameter $\theta$ is bounded by the so-called Cram\'er-Rao bound~\cite{Kay1993}
\begin{align}
\text{Var}(\theta) \geq \frac{1}{K F[M]}, \label{eq:CRBound}
\end{align} 
where $K$ is the number of measurements performed and $F[M]$ is the classical Fisher information (FI) given by 
\begin{equation}
\label{eq:fisherSP}
    F[M] = \sum_{\alpha=1}^O \text{Tr}[\rho_{\theta} M_\alpha]l_{\alpha}^2
\end{equation}
where $l_{\alpha} =\text{Tr}[\partial_{\theta}\rho_{\theta} M_\alpha]/\text{Tr}[\rho_{\theta} M_\alpha]$ are the elements of the logarithmic derivative of the probability vector, and  the trace $\text{Tr}[\cdot]$ is taken in the Hilbert space corresponding to the density operator $\rho_{\theta}$.
The Cramér-Rao bound in Eq.~\eqref{eq:CRBound},  asymptotically achievable in the limit of large $K$, shows how the FI $F[M]$ quantifies the precision attainable via a certain POVM $M$, and thus induces a hierarchy between measurements: larger values of the FI correspond to measurements yielding better estimation precision.
We finally remark that an ultimate quantum bound can be derived by defining the quantum Fisher information (QFI) as 
\begin{align}
Q = \text{Tr}[\rho_\theta L_\theta^2] \,
\end{align}
where $L_\theta$ denotes the symmetric-logarithmic derivative (SLD) operator defined via the equation $2 \partial_{\theta}\rho_{\theta} = L_\theta\rho_\theta + \rho_\theta L_\theta$.
In fact the following quantum Cram\'er-Rao bound holds~\cite{helstrom1976quantum,Holevo2011b,Paris2009}
\begin{align}
   \text{Var}(\theta) \geq \frac{1}{K F[M]} \geq \frac{1}{K Q}; \label{eq:QRBound}
\end{align}
This bound is achievable since $Q = \max_M F[M]$.

In Ref.~\cite{Menos}, the authors addressed the problem of quantifying how much the Fisher information is sensitive to small changes of the actual measurement with respect to the intended one.
The former is described by modifying the target measurement $M$, as $(1-\epsilon)M+\epsilon N$, by adding a noise component $N$, in the form of a different POVM with elements $\{N_\alpha\}$---the sum of two POVMs is meant element by element.
See Appendix B in the Supplementary Material of Ref.~\cite{Menos} for concrete examples of how to describe common measurement imperfections in this framework.
The effect a small disturbance is quantified by the relative decrease of the FI, in the limit of an infinitesimally small change:
\begin{equation}
\begin{aligned}
    \label{eq:chiSP} 
    \chi[M,N]&=\lim_{\epsilon\rightarrow 0} \frac{F[M]-F[(1-\epsilon)M+\epsilon N]}{\epsilon F[M]}\\
    &= 1+ \frac{G[N]}{F[M]},
    \end{aligned}
\end{equation}
with $G[N]=\sum_\alpha\text{Tr}[A_{\alpha} N_\alpha]$, and $A_{\alpha} = l_{\alpha}^2\rho_{\theta}-2l_{\alpha}\partial_{\theta}\rho_\theta$.
The FI MeNoS is then defined by taking the maximum of this quantity over all noise POVMs, and a closed analytical form is obtained
\begin{align}
    \sigma[M] &= \max_{N} \chi[M,N] \nonumber  \\
     &=1+\frac{1}{2F[M]}\left(l_{\alpha^{\downarrow}}^2+ l_{\alpha^{\uparrow}}^2 + \|A_{\alpha^{\downarrow}} - A_{\alpha^{\uparrow}} \|_1\right),
     \label{eq:menosSP}
\end{align}
where $l_{\alpha^{\uparrow}} = \max \left [ \{ l_\alpha | \alpha = 1, \dots,O\} \right ]  $ and $\alpha^{\uparrow}$ is the corresponding outcome; $l_{\alpha^{\downarrow}}$ and $\alpha^{\downarrow}$ are analogously defined in terms of the minimum; we recall that the trace norm of an operator is $\| X \|_1 = \Tr [ \sqrt{X X^\dag} ]$.

\section{The multiparameter FI susceptibility}
\label{s:multiMenos}

We now consider the case of a multiparameter quantum statistical model $\rho_{\bar{\theta}}$, that is a family of quantum states labeled by a vector of $P$ parameters $\bar\theta=\{\theta_1,\theta_2,...,\theta_P\}$.
In this case the Cram\'er-Rao bound is a matrix inequality for the covariance matrix 
\begin{align}
{\bf Cov}(\bar\theta) \geq \frac{1}{K} {\bf F}[M]^{-1},
\label{eq:multiCRBound}
\end{align}
i.e. the left-hand side minus the right-hand side is a positive semi-definite matrix.
The FI is now a matrix ${\bf F}[M]$ with elements 
\begin{equation}
F_{jk}[M]=\sum_\alpha \text{Tr}[\rho_{\bar\theta}M_\alpha]l_{\alpha,j}l_{\alpha,k},
\end{equation}
Here, we have introduced
\begin{equation}
l_{\alpha,j} =\frac{\text{Tr}[\partial_j\rho_{\bar\theta} M_\alpha]}{\text{Tr}[\rho_{\bar\theta} M_\alpha]}.
\label{eq:elle}
\end{equation}
with the short-hand notation $\partial_j=\partial_{\theta_j}$. Also in the multiparameter case one can define a more fundamental quantum bound via the matrix inequality
\begin{align}
{\bf Cov}(\bar\theta) \geq \frac{1}{K} {\bf F}[M]^{-1} \geq \frac{1}{K} {\bf Q}^{-1} \,,
\label{eq:multiQCRBound}
\end{align}
that can be translated to a scalar bound as
\begin{align}
\sum_j {\rm Var}(\theta_j) \geq \frac{1}{K} \text{tr}[{\bf F}[M]^{-1}] \geq \frac{1}{K}
\text{tr}[{\bf Q}^{-1}] \,.
\label{eq:multiQCRBoundscalar}
\end{align}
Note that in the formulas above, $\text{tr}[{\bf O}]$ denotes the trace taken over the matrix space ${\bf O}$, in contrast to the trace $\text{Tr}$ operating over the Hilbert space.
The elements of the QFI matrix ${\bf Q}$  are defined via the SLD operators $L_j$ for each individual parameter as $Q_{jk} = \text{Tr}[\rho_{\bar\theta}(L_j L_k + L_k L_j)/2]$.
Unlike the single-parameter case, neither the matrix bound \eqref{eq:multiQCRBound} nor the scalar bound~\eqref{eq:multiQCRBoundscalar} are achievable in general, and tighter bounds can be derived (see, e.g., Ref.~\cite{MultiPerspective} for more details).

\subsection{Scalar susceptibility for multiple parameters}

Dealing with multiple parameters, we may introduce a matrix susceptibility at fixed noise $N$, extending the definition \eqref{eq:chiSP} for the single-parameter case:
\begin{equation}
\label{eq:limit}
    {\boldsymbol \Xi}[M,N] ={\bf F}[M]^{-1} \lim_{\epsilon\rightarrow 0} \frac{{\bf F}[M]-{\bf F}[(1-\epsilon)M+\epsilon N]}{\epsilon}.
\end{equation}
The explicit evaluation of the limit results in the expression
\begin{equation}
\label{eq:chimatrix}
     {\boldsymbol \Xi}[M,N] ={\bf I}+{\bf F}^{-1}[M] {\bf G}[N],
\end{equation}
with
\begin{equation}
\begin{aligned}
    &{\bf G}[N] = \sum_\alpha \text{Tr}[{\bf A}_\alpha N_\alpha],\\
    &A_{\alpha; j,k} = l_{\alpha,j} l_{\alpha,k}\rho_{\theta} -l_{\alpha,j}\partial_k\rho_\theta-l_{\alpha,k}\partial_j\rho_\theta.
\end{aligned}
\label{eq:gea}
\end{equation}
The derivation is found in Appendix~\ref{app:EqLimit}.
However ${\boldsymbol \Xi}[M,N]$ is not a univocal matrix-valued generalization of the single-parameter scalar quantity $\chi[M,N]$: ${\bf F}[M]^{-1}$ and ${\bf G}[N]$ may not commute, thus their ordering matters.
To avoid this ambiguity and also to be able to perform an optimization over noise POVMs, we introduce a scalar quantity also in the multiparameter scenario:
\begin{align}
    X[M,N] &= \tr \left[ {\boldsymbol \Xi}[M,N] \right]  = P +\text{tr}[{\bf F}[M]^{-1}{\bf G}[N]]     \label{eq:chi}
\end{align}
Note that this resolves the ambiguity mentioned above, since the cyclicity of the trace ensures the same figure of merit is obtained for any matrix of the form $ {\bf F}[M]^{-q} {\bf G}[N] {\bf F}[M]^{-p}$ with $p+q = 1$.

A useful property of $X[N,M]$ is its invariance under reparametrizations of the statistical model.
We recall that for a new  set of parameters $\vec \varphi (\theta )=\{\varphi_1,\varphi_2,...,\varphi_P\}$ the matrix $J_{ij}= \partial_{\varphi_i}\theta_j(\vec \varphi)$, i.e. the transpose of the Jacobian matrix, links the original and reparametrised FI matrices as~\cite{Paris2009}
\begin{equation}
    \tilde{\bf{F}} = \bf{J}\,\bf{F} \, \bf{J}^{T};
\end{equation}
where we use the tilde to denote the new parameterization, also for partial derivatives $\tilde \partial_{ j}=\partial_{\varphi_j}$
and similarly for $\tilde l_{\alpha,j}$ and $\tilde A_{\alpha;j,k}$.

Since $\tilde \nabla \rho =\bf{J}\cdot\nabla \rho$, due to the definitions \eqref{eq:elle} and \eqref{eq:gea}, we have that the matrix $\mathbf{G}[ N ]$ transforms as the FI matrix
\begin{equation}
    \tilde{\bf{G}}[N]= {\bf J} \, {\bf G}[N] \, \bf{J}^{T},
\end{equation}
and thus the trace in~\eqref{eq:chi} is invariant under reparametrization, as claimed.

\subsubsection{Semidefinite program}

The FI MaS MeNoS can then be defined as in Eq.~\eqref{eq:menosSP} by maximizing over all the possible noise POVM $N$, 
\begin{align}
\Sigma[M] = \max_N X[M,N] \,.
\label{eq:menosMPdefinition}
\end{align}
This quantity then captures the robustness of the measurement $M$ for the multiparameter estimation task as a whole.
Mathematically, the function to maximize is linear in the POVM elements $N_\alpha$, which belong to the convex cone of positive semidefinite matrices $N_\alpha \geq 0$ and have to satisfy the linear constraint $\sum_{\alpha} N_\alpha = \id $.
This is thus a semidefinite program (SDP)~\cite{Boyd}, a class of optimization problems that is pivotal in quantum information science~\cite{Watrous2017,Renes2022,Skrzypczyk2023}.
From a practical point of view, many efficient algorithms for SDPs exist, which allow to obtain precise numerical solutions even for problems with relatively big matrices.
The closed-form solution in Eq.~\eqref{eq:menosSP} can be obtained for single-parameter estimation only because it is possible to reduce the optimization to a two-outcome POVM~\cite{Menos}; this is generally not the case for multiparameter problems.
This suggests an analogy with minimal-error state discrimination: for two alternative hypotheses, an expression can be found in terms of trace norm, while no close form is available more than two.
In the following we show how to simplify the formulation of the optimization problem, which is useful both for a practical implementation of the SDP and to derive upper and lower bounds on $\Sigma[M]$.

The invariance of $X[N,M]$ allows us to work in the diagonal parametrization in which the FI matrix is diagonal, i.e. reparametrizing the model such that the rows of $\mathbf{J}$ are the eigenvectors of $\mathbf{F}$.
This also implies that $\mathbf{J}$ will be orthogonal, but we could also go to the ``natural parametrization'' in which $\mathbf{F} = \id_P$, meaning that $\mathbf{J}$ is not orthogonal and includes all the $\tilde{F}_{jj}$ factors that appear below.
In the rest of the manuscript, the tilde will denote the diagonal parametrization exclusively.
This allows to express the optimization in a simpler form
\begin{align}
        \label{eq:menosMP}
        &\Sigma[M] = P +\max_{N} \sum_{\alpha=1}^{O}\sum_{j=1}^{P} \frac{\text{Tr}[\tilde A_{\alpha;jj}N_\alpha]}{\tilde  F_{jj}}\\
        & = P +\max_{N} \sum_{\alpha=1}^{O}\text{Tr}\left[ N_\alpha \left( \sum_{j=1}^{P}  \frac{ \tilde A_{\alpha;jj}}{\tilde  F_{jj}} \right) \right]. \nonumber
\end{align}
Note that one can also implement the linear constraint explicitly and eliminate one matrix variable, by setting $N_O = \id - \sum_{\alpha=1}^{O-1} N_\alpha$ and optimizing $\{N_\alpha\}_{\alpha=1,\dots,O-1}$ with the inequality constraint $\sum_{\alpha=1}^{O-1} N_\alpha \leq \id$. This optimization can be easily fed to a modelling language for a numerical evaluation of the FI MaS MeNoS.

\subsubsection{Possible reduction of the number of outcomes of $N$}

While it is not generally possible to restrict the optimization to POVMs with only two outcomes, it may still be possible to reduce the number of relevant outcomes also in the multiparameter case.
First, we introduce the functions
\begin{equation}
    f_\alpha(\vec x)=c_\alpha \| \vec x \|^2-\vec x\cdot \vec \delta_\alpha,
\end{equation}
where $c_\alpha =\text{Tr}[\rho_{\bar\theta} N_\alpha]\geq0$, $\delta_{\alpha,j} = 2\text{Tr}[\tilde \partial_j \rho_{\bar\theta} N_\alpha]/\sqrt{\tilde F_{jj}}$ and $\| \vec{x} \|^2 = \sum_{j} x_j^2$ is the Euclidean norm, making $f_\alpha$ a convex function of $\vec{x}$.
Second, we introduce the collection of $O$
vectors in $\mathbb{R}^P$:  $\vec L_\alpha = \left[ \tilde l_{\alpha,1}/\sqrt{\tilde F_{11}}, \tilde l_{\alpha,2}/\sqrt{\tilde F_{22}}, ..., \tilde l_{\alpha,P}/\sqrt{\tilde F_{PP}} \right]$, where $\alpha \in \{ 1, \dots , O \}$ labels the outcomes of $M$.
With these definitions we obtain
\begin{equation}
\label{eq:Xasasum}
    X[M,N]=P+\sum_{\alpha=1}^O f_\alpha(\vec L_\alpha).
\end{equation}

Geometrically, the vectors $\{ \vec L_\alpha \}_{\alpha=1,\dots,O}$ represent $O$ points in a $P$-dimensional Euclidean space.
The convex hull of these points forms a convex polytope and its vertices, or extremal points, must be a subset of the original points.
We denote the number of vertices with $E$ and we rearrange the outcomes of $M$ such that the vertices correspond to the first $E$ elements $\vec L_1,...,\vec L_E$.

These geometric considerations can be used to restrict the class of noise POVMs over which the maximisation of $X[M,N]$ must be carried out.
Due to the convexity of the functions $f_\alpha(\vec{x})$, they will present a maximum in correspondence of at least one of the vertices of the polytope.
Suppose that $f_{E+1}(\vec x)$ is maximised for $\vec x = \vec L_1$, and consider a generic noise POVM $N=\{N_1,N_2,...,N_{E},N_{E+1},...\}$, as well as its variation $N'=\{N_1+N_{E+1},N_2,...,N_{E},0,...\}$: by virtue of \eqref{eq:Xasasum}, we get $X[M,N']-X[M,N]=f_{E+1}(\vec L_1)-f_{E+1}(\vec L_{E+1})\geq0$.
Applying this argument to all the non-extremal outcomes we conclude that the optimal POVM $N$ that maximizes $X[M,N]$ can be restricted to have at most $E$ outcomes corresponding to the extremal points.
This transposes a property of the single-parameter FI MeNoS to the multiparameter case.
These geometrical considerations may help restricting the search of the optimal POVM $N$ for a moderate number of parameters; in fact, for arbitrary dimension finding the vertex points is not a trivial computational task.

\subsection{Upper and lower bounds}

\subsubsection{Lower bound}
The FI MaS MeNoS is, by definition, limited from below by any $X[M,N]$ at fixed $N$.
We can then choose a noise POVM with only two outcomes, corresponding to the outcomes $\alpha'$ and $\alpha''$ of $M$, which we denote as $N^{(\alpha',\alpha'')} = \{\dots, H , \dots , \id - H, \dots \}$ with $ 0 \leq H \leq \id$, where $H$ appears in the $\alpha'$-th position and $\id - H$ in the $\alpha''$-th.
This choice leads to a lower bound $\Sigma[M] \geq \Sigma_L^{(\alpha',\alpha'')}[M] $ that depends on the choice of $\alpha'$ and $\alpha''$:

\begin{align}
    & \Sigma_L^{(\alpha',\alpha'')}[M] - P =  \max_{ 0 \leq H \leq \id }  \text{tr}\left[{\bf F}[M]^{-1}{\bf G}[N^{(\alpha',\alpha'')}] \right] \\ 
    & = \|\vec L_{\alpha'}\|^2
     + \max_{ 0 \leq H \leq \id }  \text{Tr} \left[ H \left( \sum_{j=1}^P \frac{ \tilde A_{\alpha';jj}-\tilde A_{\alpha'';jj}}{\tilde  F_{jj}}  \right) \right]    \\ 
    & = \frac{\|\vec L_{\alpha'}\|^2+\|\vec L_{\alpha''}\|^2}{2} +\left\| \sum_{j=1}^P  \frac{\tilde A_{\alpha',jj}-\tilde A_{\alpha'',jj}}{2\tilde F_{jj}} \right\|_1 \,.
\end{align}
The second line is obtained by using the diagonal parametrization, as in~\eqref{eq:menosMP}, together with the form of the two-outcome POVM and the fact that $\text{Tr}\left[  \sum_{j=1}^{P}  { \tilde A_{\alpha;jj}}/{\tilde  F_{jj}}\right] = \| \vec{L}_\alpha \|^2$, since $\Tr [ \partial_j \rho_{\theta}]=0$.
From here the derivation is analogous to the single-parameter case \eqref{eq:menosSP}: the third line is a well-known identity, see e.g.~\cite[pp.~126-127]{Renes2022}, which forms the basis of the Holevo-Helstrom theorem~\cite[Theorem 3.4, p.~128]{Watrous2017}.

Finally, the tightest of this class of lower bounds, denoted by $\Sigma_L[M]$, is obtained by taking the maximum over the choice of outcome pairs $\alpha'$ and $\alpha''$:
\begin{align}
\label{eq:lbound}
\Sigma[M] &\geq \Sigma_L[M] = \max_{(\alpha',\alpha'')}\Sigma_L^{(\alpha',\alpha'')}[M].
\end{align}

\subsubsection{Upper bound}

A simple upper bound can be obtained by maximizing each term in the sum in Eq.~\eqref{eq:menosMP} separately instead of optimizing the whole sum simultaneously.
Each individual optimiziation clearly corresponds to the single-parameter FI MeNoS for each parameter in the diagonal parametrization.
Substituting the closed-form expression~\eqref{eq:menosSP}, the bound reads
\begin{align}
\Sigma[M] &\leq \Sigma^U[M] =\sum_j \sigma_j[M] \\ \label{eq:ubound}
\sigma_j[M] &=1+\frac{1}{2\tilde F_{jj}[M]}\left(\tilde l_{\alpha^{\downarrow}_j}^2+\tilde l_{\alpha^{\uparrow}_j}^2+\|\tilde A_{n_j,jj}-\tilde A_{m_j,jj}\|_1\right) \nonumber
\end{align}
where indexes $\alpha^{\downarrow}_j$ and $\alpha^{\uparrow}_j$, associated to the minimum and maximum of the set $ \{ l_{\alpha,j} \}$ respectively, may now differ depending on the parameter.

\section{Application to multiparameter quantum estimation problems}
\label{s:examples}
We will now address the evaluation of the FI MaS MeNoS $\Sigma[M]$ for measurement strategies in two paradigmatic examples of multiparameter quantum estimation. We will first consider the problem of estimating phase and phase-diffusion via qubit probes~\cite{knysh2013estimation,Vidrighin14,Altorio2015,Szczykulska_2017,Roccia_2018}, and we will then approach the estimation of centroid, separation, and relative intensities of two incoherent point sources~\cite{Rehacek2017,Rehacek18PRA,Ansari2021}.
A companion notebook with the code used to produce the numerical results for both examples is available online~\cite{AlbarelliGitHub2024}.
 
\subsection{Simultaneous estimation of phase and dephasing}
We now consider the joint estimation of a phase $\phi$ and dephasing $\Delta$ encoded in a qubit quantum statistical model $\rho_{\bar\theta}$, which in the $\sigma_z$-basis reads
\begin{equation}
    \rho_{\bar\theta} =\frac{1}{2}
    \begin{pmatrix}
         1&e^{-i\phi-\Delta}  \\
         e^{i\phi-\Delta}&1 
    \end{pmatrix}, \,\,\,\bar\theta = (\phi,\Delta)\,.
    \label{rhoTP}
    \end{equation}
This has served as a common test for multiparameter estimation~\cite{knysh2013estimation,Vidrighin14,Altorio2015,Szczykulska_2017,Roccia_2018}.  

\begin{figure}
    \centering
    \includegraphics{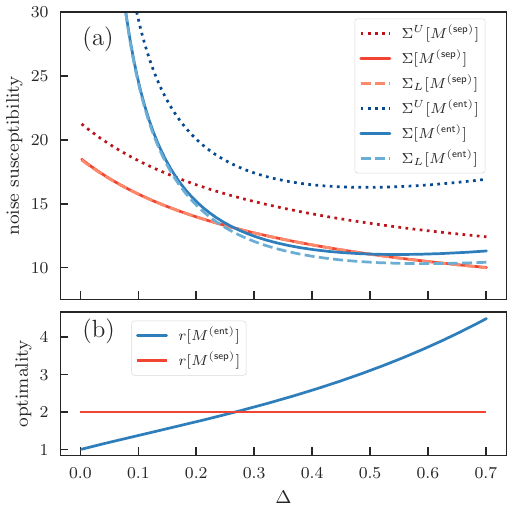}
    \caption{Panel (a): noise susceptibility (computed via SDP, solid lines), upper bound (dotted lines) and lower bound (dashed lines) for phase and phase diffusion estimation as a function of the phase diffusion parameter $\Delta$, for $\phi =\pi/4$.
    Panel (b): degree of optimality of the POVM.
    Shades of red refer to the Bell-state measurement $M^{\mathsf{(ent)}}$, while shades of blue to the separable measurement $M^{\mathsf{(sep)}}$. }
    \label{fig:phiDelta_varyingDelta}
\end{figure}

Our aim is to quantify the FI MaS MeNoS associated to the two different strategies: i) a separable-measurement strategy, corresponding to a four-outcome single-qubit POVM $M^{\sf (sep)}$ such that
\begin{align}
M^{\sf (sep)}_1 &= |+_x\rangle \langle +_x|/2  \,\,\,\, &M^{\sf (sep)}_2 &= |-_x\rangle \langle -_x|/2 \nonumber \\
M^{\sf (sep)}_3 &= |+_y\rangle \langle +_y|/2  \,\,\,\, &M^{\sf (sep)}_4 &= |-_y\rangle \langle -_y|/2 \,
\end{align}
where $|\pm_x\rangle$ and $|\pm_y\rangle$ are eigenstates respectively of $\sigma_x$ and $\sigma_y$; 
ii) an entangled-measurement strategy on two copies $\rho_{\bar\theta}\otimes \rho_{\bar\theta}$, corresponding to the Bell measurement POVM
\begin{align}
M^{\sf (ent)}_1 &= |\Psi_+\rangle \langle \Psi_+|  \,\,\,\, &M^{\sf (ent)}_2 &= |\Psi_-\rangle \langle \Psi_-| \nonumber \\
M^{\sf (ent)}_3 &= |\Phi_+\rangle \langle \Phi_+|  \,\,\,\, &M^{\sf (ent)}_4 &= |\Phi_-\rangle \langle \Phi_-| \,
\end{align}
where $|\Psi_\pm\rangle$ and $|\Phi_\pm\rangle$ denote the four Bell states.
It has been in fact discussed how the use of a Bell-measurement on two copies of the state \eqref{rhoTP} can result in an improved extraction of the FI with respect to separable measurements~\cite{Roccia_2018}.
In particular, following the idea put forward in Ref.~\cite{Belliardo2021}, one can define a parameter quantifying how the estimation via a certain POVM $M$ acting on $m$ copies of the state $\rho_{\bar\theta}$ is far from the multiparameter bound as 
\begin{align}
r[M] = \frac{m \, \text{tr}[{\bf F}[M]^{-1}]}{\text{tr}[{\bf Q}^{-1}]}\geq 1 \,.
\label{eq:rBelliardo}
\end{align}
The lower bound $r[M] = 1$ is achieved whenever the scalar bound \eqref{eq:multiQCRBoundscalar} is  saturated.
One can prove that, given the quantum statistical model $\rho_{\bar\theta}$ in Eq.~\eqref{rhoTP}, the two SLD operators satisfy the weak-commutativity condition $\Tr[\rho_{\bar\theta}[L_\phi,L_\Delta]] = 0$.
This implies that the quantum statistical model is asymptotically classical~\cite{MultiPerspective}, meaning that a POVM $M$ acting on an asymptotically large number of copies $m\to \infty$ exists such that the bound \eqref{eq:multiQCRBoundscalar} is achievable, and thus $r[M] \to 1$ in this limit.
By considering the two strategies introduced above, one has
\begin{align}
    r[M^{\sf (sep)}] &=2 \,, \\
    r[M^{\sf (ent)}] &=\frac{1-2 \,e^{4 \Delta}}{1-2\,e^{2 \Delta}} \,, \,\,\,\text{for} \,\, \phi=\pi/4 \,
\end{align}
where $r[M^{\sf (ent)}]$ has been minimized over the phase parameter $\phi$.
One can thus show that for a specific value of the phase $\phi$ and for vanishing dephasing $\Delta\approx 0$ the scalar Cram\'er-Rao bound \eqref{eq:multiQCRBoundscalar} is already saturated by implementing a two-copy entangled measurement strategy, while for larger values of dephasing $\Delta\gtrsim 0.27$ the Bell measurement performs worse than the separable one, see panel (b) of Fig.~\ref{fig:phiDelta_varyingDelta}.

We study the behaviour of the FI MaS MeNoS for the two POVMs in this regime, in order to understand any possible relationship between efficiency in the multiparameter estimation and noise susceptibility.
In Fig.~\ref{fig:phiDelta_varyingDelta} we plot the FI MaS MeNoS $\Sigma[M]$, that has been obtained by a numerical evaluation using Eq.~\eqref{eq:menosMP} as a semidefinite program (SDP) in Python using CVXPY~\cite{Diamond2016}.
For comparison, we also report the corresponding upper bound $\Sigma^U[M]$ and lower bound $\Sigma_L[M]$ both for the separable measurement $M^{\sf (sep)}$ and for the entangled measurement $M^{\sf (ent)}$.
We fix $\phi=\pi/4$ and plot the quantities as a function of $\Delta$.

First, we observe that the bounds $\Sigma_L$ and $\Sigma^U$ give reliable information on the actual multiparameter susceptibilities.
On the one hand, we observe that for this problem the exact result obtained by solving the SDP is close to the lower bound $\Sigma_L$, being almost identical in the case of separable measurements on a single copy.
On the other hand, the bounds capture the behaviour in the limit of vanishing noise $\Delta \to 0$, where the susceptibility $\Sigma[M^{\sf (ent)}]$ diverges, while $\Sigma[M^{\sf (sep)}]$ tends to a finite value.
Thus, even if a projection onto Bell states allows to saturate the multiparameter quantum Cramér-Rao bound as $\Delta \to 0$, such a measurement becomes highly susceptible to imperfections in its implementation. This supports qualitative observations that had been put forward in Ref.~\cite{Roccia_2018}.
Intriguingly, the parameter region $\Delta \lesssim 0.27$, where the entangled measurement performs better, corresponds also to the region in which its FI MaS MeNos becomes larger than for the separable POVM.
For intermediate values of $\Delta$ there is a small region where the susceptibility of the entangled measurement sits slightly below the separable case, but as $\Delta$ increases $\Sigma[M^{\sf (ent)}]$ starts growing again, while $\Sigma[M^{\sf (sep)}]$ decreases monotonically.

\subsection{Multiparameter quantum metrology of weak incoherent point sources}
As our second example, we consider the problem of estimating the separation $s$ of two incoherent sources.
These are found close to their centroid position $x_c$, with a relative intensity $q$.
All these parameters are unknown and must be estimated at once.
Under the assumption of weak sources, a suitable quantum statistical model to describe the state of a single photon at the image plane is given by
\begin{align}
\rho_{\bar\theta} = q |\psi_+\rangle\langle\psi_+| + (1-q) |\psi_-\rangle\langle\psi_-|
\label{eq:qsm_2}
\end{align}
where the states $|\psi_\pm\rangle$ are defined by projecting on the $x$-basis as $x$-displaced Gaussian point spread functions 
\begin{align}
\langle x | \psi_\pm\rangle = g(x,x_c \pm s/2)
\end{align}
with
\begin{align}
g(x,x_0) = \frac{1}{(2\pi)^{1/4}}\exp\left\{-\frac{(x-x_0)^2}{4}\right\}.
\end{align}

A proper analysis on the ultimate quantum limit on this kind of estimation has been extensively discussed~\cite{Hradil2017,Rehacek18PRA}, showing how the Rayleigh criterion can be in principle overcome also in the multiparameter scenario.
While the most immediate interpretation is given in terms of optical sources, the same formalism can be readily applied to a mixture of incoherent short pulses with a given time-delay $s$, and an experimental verification of quantum timing resolution has been shown in Ref.~\cite{Ansari2021}.

The optimal measurement that has been realized in Ref.~\cite{Ansari2021}, was previously theoretically investigated in Refs.~\cite{Rehacek2017,Rehacek18PRA}.
For small values of $s$ this consists in a $5$-outcome POVM with $M_j = |v_j\rangle \langle v_j|$, $j=0,\dots,3$, and $M_4 = \mathbbm{1} - \sum_{j=0}^3 M_j$. In particular one shows that 
\begin{align}
|v_j\rangle = \sum_{k=0}^3 w_{j,k} |\Phi_k\rangle \label{eq:optimalmeas}
\end{align}
where the states $|\Phi_k\rangle$ are defined by their projection on the $x$-basis as 
\begin{align}
    \langle x| \Phi_n\rangle =g(x,\bar{x})H_n \left(\frac{x-\bar{x}}{\sqrt{2}} \right),
\end{align}
where $H_n(x)$ is the $n$-th Hermite-Gauss polynomial: $H_n(x)=(-1)^n e^{x^2}(d^n e^{-x^2} /dx^n)$. These states thus represent Hermite-Gauss modes centered on the position $\bar x$.
The weight matrix entering in \eqref{eq:optimalmeas} is
\begin{equation}
    {\bf w}= \begin{pmatrix}
       0&\frac{1}{\sqrt{6}}&\frac{1}{\sqrt{2}}&-\frac{1}{\sqrt{3}}\\
       0&\frac{1}{\sqrt{6}}&-\frac{1}{\sqrt{2}}&-\frac{1}{\sqrt{3}}\\
       \sqrt{\frac{2}{5}}&\sqrt{\frac{2}{5}}&0&\frac{1}{\sqrt{5}}\\
-\sqrt{\frac{3}{5}}&\frac{2}{\sqrt{15}}&0&\sqrt{\frac{2}{15}}
\end{pmatrix}\,,
\end{equation}
while the central position $\bar{x}$ for the optimal measurement is equal to $\bar{x}_{\rm opt}=x_c+(q-1/2)s$~\cite{Rehacek2017}.
Contrary to the direct intensity detection, this measurement strategy shows a non-vanishing precision also in the limit $s\rightarrow 0$, when $q = 1/2$.

\begin{figure}
\includegraphics{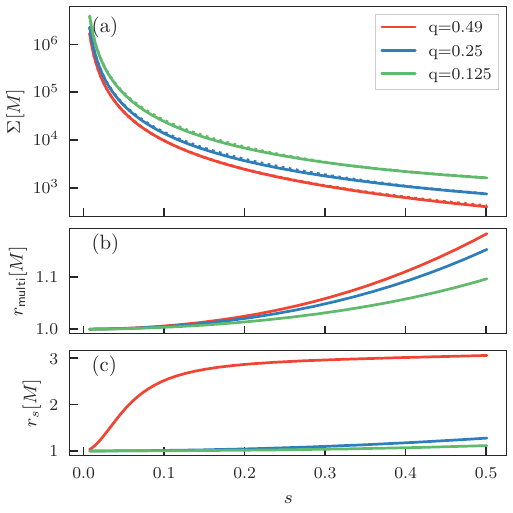}
 
    \caption{
    Panel (a): Noise susceptibility $\Sigma[M]$ (solid lines) and upper bound (dashed lines) for the estimation of the parameters $(s,x_c,\theta)$ of incoherent point sources, as a function of the spatial separation $s$.
    Panel (b): optimality parameter $r_{s}[M]$ for the estimation of $s$ as a single parameter.
    Panel (c): optimality parameter $r_{\sf multi}[M]$ for the joint estimation of the three parameters.
    }
    \label{f:incoherent}
\end{figure}

In order to show the optimality of the POVM in the small $s$ regime we consider the quantities
\begin{align}
r_{s}[M] = \frac{({\bf F}[M]^{-1})_{1,1}}{({\bf Q}^{-1})_{1,1}} \geq 1 \,,
\label{eq:belli1}
\end{align}
and 
\begin{align}
r_{\sf multi}[M] = \frac{\hbox{tr}[{\bf F}[M]^{-1}]}{\hbox{tr}[{\bf Q}^{-1}]} \geq 1\,.
\label{eq:belli2}
\end{align}
The numerator and denominator of the first quantity $r_{s}[M]$ correspond to the classical and quantum Cramér-Rao bound for the estimation of the sources separation $s$, taking into account the centroid $x_c$ and the relative intensity $q$ as (unknown) nuisance parameters~\cite{Suzuki_2020,Tsang2019}. 
This means that, while the parameter of interest is only $s$, we do not neglect that estimates of $s$ are generally correlated with the estimates of the other two parameters.
This increases the error, since $({\bf F}[M]^{-1})_{1,1} \geq 1 / {\bf F}[M]_{1,1} $ (and analogously for the QFI), where the right-hand side is the single-parameter Cramér-Rao bound (in which $q$ and $x_c$ are assumed to be perfectly known).

The second quantity in Eq.~\eqref{eq:belli2}, analogous to the one defined for the previous example in Eq.~\eqref{eq:rBelliardo}, quantifies the optimality of the measurement strategy for the estimation of all the parameters (in this scenario the measurement is applied to a single copy of the quantum state, i.e. $m=1$).
Both quantities \eqref{eq:belli1}  and \eqref{eq:belli2} achieve their minimum value when the measurement is optimal, i.e. it saturates the (single- or multi-parameter) quantum Cram\'er-Rao bound.
As we can notice from panels (b) and (c) of Fig.~\ref{f:incoherent}, this measurement enjoys such an optimality when the spatial separation $s$ goes to zero, both for the estimation of $s$ only and of all three parameters.
The quality of the estimation of both $s$, as captured by \eqref{eq:belli1}, and of all three parameters, captured by \eqref{eq:belli2}, increases as the intensity unbalance $q$ increases, {\it i.e.} as $q$ moves away from the balanced $q=1/2$ towards the values $q=0$ (or equivalently $q=1$).
However, we see a qualtitive difference between the estimation of $s$ and multiparameter estimation, the latter performing much closer to optimality, $r_{\sf multi}[M]\approx 1$. 
We remind, however, that this metric should be adopted to compare the precision obtainable with a given measurement to the relative Cram\'er-Rao bound, but not to compare different estimation problems.
For example, the absolute error on the estimation of $s$ is independent of its value only for $q=1/2$; in other cases it grows as $s\to 0$, but more favourably compared to direct intensity measurements~\cite{Rehacek18PRA}.

The fact that multiparameter estimation performs closer to optimality than the single parameter case may appear counterintuitive.
We can partly explain this behaviour by noticing that the uncertainties about the different parameters may vary by orders of magnitude, even when each one reaches its individual Cram\'er-Rao bound.
Consequently, if equal weights are applied, the sum of the variances can be dominated by the contribution of one parameter over the others---specifically, the uncertainty on $q$ is by far the leading term.
Since the measurement is optimal for the estimation of $q$, we get $r_{\sf multi}[M]\approx 1$ due to this dominating behaviour regardless the behaviour of the other two contributions.

It is possible to evaluate the multiparameter susceptibility and its upper and lower bounds using the expressions in Sec.~\ref{s:multiMenos}, even if the states $\ket{\psi_{\pm}}$ are infinite-dimensional.
As a matter of fact, the finite rank of the quantum statistical model makes it possible to work with finite-dimensional matrices for $\rho_{\bar{\theta}}$ and its derivatives, by introducing a suitable orthonormal basis of $\mathrm{span}\left\{ \ket{\psi_{+}},  \ket{\psi_{-}},  \ket{\partial_1 \psi_{+}},  \ket{\partial_1 \psi_{-}}\right\}$ (derivatives with respect to the other two parameters still give operators supported on this subspace).
Details on these calculations are expanded in Appendix~\ref{app:incoherent}.
After this procedure, we can adopt the same numerical techniques as for the previous example; the code is available at~\cite{AlbarelliGitHub2024}.

In panel (a) of Fig.~\ref{f:incoherent} we show the multiparameter susceptibility $\Sigma[M]$ as a function of $s$ for different values of $q$.
Since we are considering the optimal measurement centered around $\bar{x}_{\rm opt}$, the problem enjoys a translational invariance and the figures of merit do not depend on the specific value of $x_c$.
The exact value of $\Sigma[M]$ obtained with the SDP is close to the lower bound (indistinguishable in the plot) also for this model.
While the discrepancy with the upper bound is more significant, it is still a small difference relative to the large values that the quantities attain.
For this reason the dashed lines corresponding to the upper bound are barely appreciable in logarithmic scale.

We observe that both the noise susceptibility $\Sigma[M]$ and the optimality figures of merit $r_{s}[M]$ and $r_{\sf multi}[M]$ are monotonic in $q$; the former is decreasing while the latter is increasing (clearly, the situation would be reversed for $q>1/2$). 
Thus, we can conclude that when the measurement gets more optimal the price to pay is an increased sensitivity to noise (a similar conclusion was drawn in previous example, albeit only in the region $\Delta \lesssim 0.4$ in which entangled measurements perform well).
More importantly, we also observe that the susceptibility $\Sigma[M]$ is diverging for $s \to 0$: i.e. when the measurement is optimal, small errors in its implementation of the POVM are highly detrimental.

\section{Conclusions and outlooks}

Local quantum estimation theory has provided an instrumental set of tools for the theoretical analysis of quantum metrology.
However, to properly address practical implementations, we argue that the performance of a particular measurement (or more generally a particular metrological strategy) should be assessed beyond its ability to saturate the fundamental quantum Cramér-Rao bound.
Following Ref.~\cite{Kurdzialek2023a}, we have approached this issue from the observation that all measurements are necessarily imperfect, extending the framework to multiple parameters.

The analysis of multiparameter measurements is beset on all sides by the noncommutativity of the individual optimal measurements, as well as by statistical correlations.
We have nevertheless succeeded in finding a workable definition for the susceptibility to measurement noise; this is able to provide concise information in the form of a scalar figure of merit, which is efficiently computable as an SDP.
We have also established closed-form upper and lower bounds; our examples show these can be sufficiently tight to be useful.

The examples also revealed that better performance of a measurement, in the sense of being closer to attaining the quantum Cramér-Rao bound, often comes attached to an increased sensitivity to noise.

We observe that in Ref.~\cite{Vasilyev2024} a new cost function for multiparameter estimation has been introduced. This takes into account the idea that a measurement should perform well not only at the true parameter value but also in its vicinity.
We see this as a complementary approach to ours: both a purely local analysis and the assumption of perfect measurements are idealizations. Going beyond them, new ways to assess the performance of a measurement for parameter estimation are introduced that are different from the evaluation of the corresponding Cramér-Rao bound.
Intriguingly, the optimal measurement for noiseless single-parameter metrology presented in Ref.~\cite{Vasilyev2024} corresponds to the measurement that minimize the FI noise susceptibility, found in Ref.~\cite{Kurdzialek2023a}.
It is an interesting open question to understand if deeper connections between these two frameworks may exist, especially in the more intricate case of multiple parameters.

In conclusion, our work enriches the toolkit for the inspection of metrological schemes, being applicable to classical and quantum schemes equally well.
Future developments will consider Bayesian schemes, which may affect robustness through prior knowledge, as well as different weighting among the parameters.

\section*{Acknowledgements}
We thank R. Demkowicz-Dobrza\'nski for discussions.\\
MB and IG acknowledge support from the EU Commission (H2020 FET-OPEN-RIA STORMYTUNE, Grant Agreement No. 899587), and from MUR (PRIN22-RISQUE-2022T25TR3).
MGG acknowledges support from from Italian Ministry of Research and Next Generation EU via the
PRIN 2022 project CONTRABASS (contract n.2022KB2JJM).
FA acknowledges support from Marie Skłodowska-Curie Action EUHORIZON-MSCA-2021PF-01 (project QECANM, grant n. 701154).
\newpage
\appendix
\section{On the evaluation of Eq.~\eqref{eq:limit}}
\label{app:EqLimit}
The limit in Eq.~\eqref{eq:limit} can be calculated from the elements of the $\bf F$ matrix:
\begin{equation}
\begin{aligned}
&F_{j,k}[(1-\epsilon)M+\epsilon N] =\\
&=\sum_\alpha \Tr[\rho_{\bar \theta}M_\alpha] \frac{((1-\epsilon)l_{\alpha,j}+\epsilon m_{\alpha,j}) ((1-\epsilon)l_{\alpha,k}+\epsilon m_{\alpha,k})}{(1-\epsilon)+\epsilon\frac{\Tr[\rho_{\bar \theta}N_\alpha]}{\Tr[\rho_{\bar \theta}M_\alpha]}},  
\end{aligned}
\end{equation}
where we have used the linearity of the trace, and have defined $m_{\alpha,j} = \Tr[\partial_j\rho_{\bar \theta}N_\alpha]/\Tr[\rho_{\bar \theta}M_\alpha]$. This implies that
\begin{widetext}
\begin{equation}
\frac{F_{j,k}[M]-F_{j,k}[(1-\epsilon)M+\epsilon N]}{\epsilon} =\frac{1}{\epsilon}\sum_\alpha \Tr[\rho_{\bar \theta}M_\alpha]l_{\alpha,j}l_{\alpha,k} -\Tr[\rho_{\bar \theta}M_\alpha] \frac{((1-\epsilon)l_{\alpha,j}+\epsilon m_{\alpha,j}) ((1-\epsilon)l_{\alpha,k}+\epsilon m_{\alpha,k})}{(1-\epsilon)+\epsilon\frac{\Tr[\rho_{\bar \theta}N_\alpha]}{\Tr[\rho_{\bar \theta}M_\alpha]}}.  
\end{equation}
\end{widetext}
For $\epsilon\simeq 0$, one can take the approximation 
\begin{equation}
    \frac{1}{(1-\epsilon)+\epsilon \frac{\Tr[\rho_{\bar \theta}N_\alpha]}{\Tr[\rho_{\bar \theta}M_\alpha]}}\simeq 1-\epsilon\left(\frac{\Tr[\rho_{\bar \theta}N_\alpha]}{\Tr[\rho_{\bar \theta}M_\alpha]}-1\right),
\end{equation}
that leads, in the limit of an infinitely small disturbance, to the expression
\begin{equation}
\begin{aligned}
&\lim_{\epsilon\rightarrow 0}\frac{F_{j,k}[M]-F_{j,k}[(1-\epsilon)M+\epsilon N]}{\epsilon}=\\
&= F_{j,k}[M] + \sum_\alpha \text{Tr}[ A_{\alpha;jk} N_\alpha],  
\end{aligned}
\end{equation}
and, consequently, to our claim.

\section{Multiparameter quantum metrology of incoherent point sources - finite dimensional calculations}
\label{app:incoherent}

In this appendix we present the details of the calculations needed to evaluate $\Sigma[M]$ for the estimation of incoherent optical point sources.
By following the idea pursued in Ref.~\cite{Menos}, we will first define an orthonormal basis of vectors 
$\{|b_j\rangle\}_{j=1}^4=\{|0\rangle_s, |1\rangle_s, |0\rangle_a, |1\rangle_a\}$ 
We need only four basis elements because the derivative with respect to $s$ and $x_c$ are linearly dependent, since $\ket{\partial_2  \psi_{\pm}} = \pm 2 \ket{\partial_1  \psi_{\pm}} $.
We can then use this basis to express the quantum state $\rho_{\bar{\theta}}$ and its derivatives $\partial_j \rho_{\bar\theta}$ with respect to the parameters $\bar\theta =(s, x_c, q)$.

We define the orthonormal basis elements as
\begin{align}
|0\rangle_s &= K_{0s}\left( |\psi_+\rangle + |\psi_-\rangle\right) \,,\\
|0\rangle_a &= K_{0a} \left( |\psi_+\rangle - |\psi_-\rangle\right) \,,\\
|1\rangle_s &= K_{1s} \left( |\partial_1 \psi_+\rangle + |\partial_1\psi_-\rangle - \xi K_{0s} |0\rangle_s \right) \,, \\
|1\rangle_a &= K_{1a} \left( |\partial_1 \psi_+\rangle - |\partial_1\psi_-\rangle + \xi K_{0a} |0\rangle_a \right) \,,
\end{align}
where 
\begin{align}
\nu &= \langle \psi_- | \psi_+\rangle =  e^{-\frac{s^2}{8}} \,\\
\xi &= \langle \psi_+ | \partial_1 \psi_-\rangle + \langle \psi_- | \partial_1 \psi_+\rangle  =  -\frac{\nu \,s}{4}\,\\
\lambda&= 2 \langle \partial_1 \psi_+ | \partial_1 \psi_-\rangle = \frac{\nu}{32} \left[s^2 - 4 \right] \, \\
K_{0s} &= \frac{1}{\sqrt{2 ( 1+\nu)}}\,, \\
K_{0a} &= \frac{1}{\sqrt{2 ( 1-\nu)}}\,, \\
K_{1s} &= \left[ \frac{1}{8} + \lambda - (K_{0s} \xi)^2 \right]^{-\frac12} \,\\
K_{1a} &= \left[ \frac{1}{8} - \lambda - (K_{0a} \xi)^2 \right]^{-\frac12}.
\end{align}
The matrix
\begin{equation}
{\mathcal K} = \frac{1}{2}\left(
\begin{array}{cccc}
 \frac{1}{K_{0s}} & \frac{1}{K_{0a}} & 0 & 0 \\
 \frac{1}{ K_{0s}} & -\frac{1}{ K_{0a}} & 0 & 0 \\
 K_{0s} \xi  & -K_{0a} \xi & \frac{1}{ K_{1s}} & \frac{1}{ K_{1a}} \\
 K_{0s} \xi  & K_{0a} \xi  & \frac{1}{ K_{1s}} & -\frac{1}{ K_{1a}} \\
\end{array}
\right)
\end{equation}
is used to express the non-orthogonal basis vectors $\{|v_j\rangle\}_{j=1}^4= \left\{ \ket{\psi_{+}}, \ket{\psi_{-}} , \ket{\partial_1 \psi_{+} },\ket{\partial_1 \psi_{-} } \right\}$ as linear combinations of the orthonormal basis as $\ket{v_i} = \sum_{k} {\mathcal K}_{ik} \ket{b_k}$.

The quantum state $\rho_{\bar\theta}$ and its derivatives can be written written as finite dimensional matrices with respect to the orthonormal basis $b$ through the matrix ${\mathcal K}$ as follows
\begin{align}
    \rho_{\bar\theta} &= {\mathcal K}^{T}
    \left( \begin{array}{cccc}
q & 0 & 0 & 0 \\
 0& 1-q & 0 & 0 \\
 0  & 0 & 0 & 0 \\
 0  & 0 & 0 & 0  \\
\end{array}
\right) {\mathcal K} \\
        \partial_1 \rho_{\bar\theta} &= {\mathcal K}^{T}
    \left( \begin{array}{cccc}
0 & 0 & q & 0 \\
 0& 0 & 0 & 1-q \\
 q & 0 & 0 & 0 \\
 0  & 1-q & 0 & 0  \\
\end{array}
\right) {\mathcal K} \\
 \partial_2 \rho_{\bar\theta} &= 2 {\mathcal K}^{T}
    \left( \begin{array}{cccc}
0 & 0 & q & 0 \\
 0& 0 & 0 & q-1 \\
  q & 0 & 0 & 0 \\
 0  & q-1 & 0 & 0  \\
\end{array}
\right) {\mathcal K} \\
\partial_3 \rho_{\bar\theta} &= {\mathcal K}^{T}
    \left( \begin{array}{cccc}
1 & 0 & 0 & 0 \\
 0& -1 & 0 & 0 \\
 0 & 0 & 0 & 0 \\
 0  & 0 & 0 & 0  \\
\end{array}
\right) {\mathcal K} \,.
\end{align}

We conclude this section by mentioning that for this problem it is easier to compute the QFI working directly with non-orthogonal bases~\cite{Genoni2019,Fiderer2021}; similarly, a formulation of the FI MaS MeNoS in terms of nonorthogonal bases is also possible.
We have not pursued this approach here to avoid unnecessary complications in the notation and to remain consistent with the treatment of Ref.~\cite{Menos}.


\nocite{apsrev41Control}
\bibliography{MPMenos}

\end{document}